\documentclass [a4paper,final,conference,10pt]{E-TEMS}
\usepackage[utf8]{inputenc}
\usepackage[english]{babel}
\usepackage{amsmath}
\usepackage{graphicx}
\usepackage{multirow}
\usepackage{cite}
\usepackage{amsmath,amssymb,amsfonts}
\usepackage{algorithmic}
\usepackage{textcomp}
\usepackage{xcolor}
\usepackage{comment}
\usepackage{array}
\usepackage{hyperref}
\usepackage{moresize}
\usepackage{listings}

\definecolor{darkgreen}{rgb}{0.0,0.4,0.0} 
\definecolor{darkred}{rgb}{0.6,0.1,0.1}
\definecolor{lightgray}{gray}{.98}
\definecolor{medgray}{gray}{.70}
\definecolor{darkgray}{gray}{.40}
\definecolor{lightviolet}{rgb}{0.7,0,0.7} 
\definecolor{lightlightviolet}{rgb}{1.0,0.7,1.0} 
\definecolor{darkviolet}{rgb}{0.5,0.1,0.5}
\definecolor{darkredviolet}{rgb}{0.6,0.1,0.4}
\definecolor{limegreen}{rgb}{0.2,0.7,0.2}
\definecolor{navyblue}{RGB}{0,0,128}
\definecolor{aquamarine}{RGB}{102,205,170}

\definecolor{strictRED}{RGB}{184,0,0}
\definecolor{specificationTURQUOISE}{RGB}{0,128,153}
\definecolor{assumptionGREEN}{RGB}{0,128,0}
\definecolor{interruptBLUE}{RGB}{0,0,128}
\definecolor{committedORCHID}{RGB}{54,22,89}
\definecolor{urgentORCHID}{RGB}{74,28,109}
\definecolor{requestedORCHID}{RGB}{104,34,139}
\definecolor{eventuallyORCHID}{RGB}{154,50,205}

\lstdefinelanguage{SMLX}
{
	basicstyle=\ssmall\ttfamily, %
	frame=single, 
	framextopmargin=0pt,
	framexbottommargin=0pt,
	framexleftmargin=0pt,
	xleftmargin=16pt,
	xrightmargin=3pt,
	morekeywords=[1]{system, domain, scenario, bind, to, 
		message, non, spontaneous, events, specification, 
		alternative, if, collaboration, role, with, dynamic, 
		bindings, or, and, null, define, as, 
		constraints, import, static, parameter, ranges, var, EInt, 
		controllable},
	morekeywords=[2]{strict},
	morekeywords=[3]{forbidden, violation},
	morekeywords=[4]{interrupt},
	morekeywords=[5]{guarantee},
	morekeywords=[6]{assumption}, 
	morekeywords=[7]{committed}, 
	morekeywords=[8]{urgent},
	morekeywords=[9]{requested},
	morekeywords=[10]{eventually},
	keywordstyle=[1]\color{darkviolet}\textbf,
	keywordstyle=[2]\color{strictRED}\textit,
	keywordstyle=[3]\color{strictRED}\textit,
	keywordstyle=[4]\color{interruptBLUE}\textit,
	keywordstyle=[5]\color{specificationTURQUOISE}\textbf,
	keywordstyle=[6]\color{assumptionGREEN}\textbf,
	keywordstyle=[7]\color{committedORCHID}\textit,
	keywordstyle=[8]\color{urgentORCHID}\textit,
	keywordstyle=[9]\color{requestedORCHID}\textit,
	keywordstyle=[10]\color{eventuallyORCHID}\textit,
	sensitive=false,
	morecomment=[l][\color{darkgreen}\textit]{//},
	morecomment=[s][\color{darkgreen}\textit]{/*}{*/}, 
	morestring=[b][\color{blue}]",
	tabsize=1,
	moredelim = [s][\color{specificationTURQUOISE}\textbf]{guarantee}{scenario},
	moredelim = [s][\color{assumptionGREEN}\textbf]{assumption}{scenario},
	backgroundcolor=\color{lightgray}
}

\lstdefinestyle{SMLXStyle} {language=SMLX}

\lstdefinelanguage{SMLConfig}
{
	basicstyle=\ssmall\ttfamily,
	frame=single, 
	framextopmargin=0pt,
	framexbottommargin=0pt,
	framexleftmargin=0pt,
	xleftmargin=16pt,
	xrightmargin=3pt,
	morekeywords=[1]{symbolic, import, configure, specification, use, 
		instancemodel, symbolic, parameters, attributes, symbolic, state, matching, 
		off, under, approximation, on, rolebindings, collaboration, object, plays, 
		role, role1},
	keywordstyle=[1]\color{darkviolet}\textbf,
	sensitive=false,
	morecomment=[l][\color{darkgreen}\textit]{//},
	morecomment=[s][\color{darkgreen}\textit]{/*}{*/}, 
	morestring=[b][\color{navyblue}\textit]",
	stringstyle=\color{navyblue},
	tabsize=1,
	backgroundcolor=\color{lightgray}
}

\lstdefinestyle{SMLConfigStyle} {language=SMLConfig}
\lstdefinelanguage{Java}
{
	basicstyle=\ssmall\ttfamily,
	frame=single, 
	framextopmargin=0pt,
	framexbottommargin=0pt,
	framexleftmargin=0pt,
	xleftmargin=16pt,
	xrightmargin=3pt,
	morekeywords=[1]{public, private, class, extends, protected, void,
		new, throws, null, if, else},
	morekeywords=[2]{STRICT},
	morekeywords=[3]{@Override},
	morekeywords=[4]{car, oc, cp}, 
	keywordstyle=[1]\color{darkviolet}\textbf,
	keywordstyle=[2]\color{javablue}\textbf,
	keywordstyle=[3]\color{darkgray},
	keywordstyle=[4]\color{navyblue},
	sensitive=false,
	morecomment=[l][\color{javagreen}\textit]{//},
	morecomment=[s][\color{javagreen}\textit]{/*}{*/}, 
	morestring=[b][\color{javablue}\textit]",
	stringstyle=\color{navyblue},
	tabsize=1,
	backgroundcolor=\color{lightgray}
}

\definecolor{javared}{rgb}{0.6,0,0} 
\definecolor{javablue}{rgb}{0,0,0.9} 
\definecolor{javagreen}{rgb}{0.25,0.5,0.35} 
\definecolor{javapurple}{rgb}{0.5,0,0.35} 
\definecolor{javadocblue}{rgb}{0.25,0.35,0.75} 

\lstdefinestyle{JavaStyle} {language=Java}

\lstdefinelanguage{Kotlin}{
	basicstyle=\ssmall\ttfamily,
	frame=single, 
	framextopmargin=0pt,
	framexbottommargin=0pt,
	framexleftmargin=0pt,
	xleftmargin=16pt,
	xrightmargin=3pt,
	comment=[l]{//},
	commentstyle={\color{darkgray}\ttfamily},
	emph={delegate, filter, first, firstOrNull, forEach, lazy, map, mapNotNull, println, return@, event, sends, request, requestParamValuesMightVary},
	emphstyle={\color{darkviolet}},
	identifierstyle=\color{black},
	numberstyle=\color{darkgreen},
	keywords=[1]{ abstract, actual, as, as?, break, by, companion, continue, data, do, dynamic, else, enum, expect, false, final, for, get, if, import, in, interface, internal, is, null, object, override, package, private, public, return, set, super, suspend, this, throw, true, try, typealias, val, var, vararg, when, where, while},
	keywordstyle=[1]{\color{javablue}\bfseries},
	keywords=[2]{@Deprecated, @JvmField, @JvmName, @JvmOverloads, @JvmStatic, @JvmSynthetic, @Test, Array, Byte, Double, Float, Int, Integer, Iterable, Long, Short, String, scenario, class, fun},
	keywordstyle=[2]{\color{javablue}},	
	keywords=[3]{interruptingEvents, forbiddenEvents, it}, 
	keywordstyle=[3]{\color{darkviolet}\bfseries},
	keywords=[4]{scenario, cycleScenario, runTest, Given, When, Then, And, But}, %
	keywordstyle=[4]{\textit},
	keywords=[5]{coolantTemp, deratingFactor}, 
	keywordstyle=[5]{\color{darkviolet}\bfseries\underbar},
	keywords=[6]{currentTemp}, 
	keywordstyle=[6]{\underbar},
	morecomment=[s]{/*}{*/},
	morecomment=[s][\color{black}]{`}{`},
	morestring=[b]",
	morestring=[s]{"""*}{*"""},
	sensitive=true,
	stringstyle={\color{javagreen}\ttfamily},
}

\lstdefinestyle{KotlinStyle} {language=Kotlin}

\newcommand{\lstinlineKotlin}[1]{\lstinline[language=Kotlin,basicstyle=\small\ttfamily]{#1}}

\lstdefinelanguage{Gherkin}{
	basicstyle=\ssmall\ttfamily,
	frame=single, 
	framextopmargin=0pt,
	framexbottommargin=0pt,
	framexleftmargin=0pt,
	xleftmargin=16pt,
	xrightmargin=3pt,
	comment=[l]{//},
	commentstyle={\color{darkgray}\ttfamily},
	emph={@software, @charging },
	emphstyle={\color{limegreen}},
	identifierstyle=\color{black},
	keywords={Given, When, Then, And},
	keywordstyle={\color{violet}\ttfamily},
	morecomment=[s]{/*}{*/},
	morestring=[b]",
	morestring=[s]{"""*}{*"""},
	ndkeywords={Scenario, Example, Feature},
	ndkeywordstyle={\color{darkviolet}\bfseries},
	sensitive=true,
	stringstyle={\color{javagreen}\ttfamily},
	numbers=none
}

\lstdefinestyle{GherkinStyle} {language=Gherkin}

\lstdefinelanguage{Jira}{
	basicstyle=\ssmall\ttfamily,
	frame=single, 
	framextopmargin=0pt,
	framexbottommargin=0pt,
	framexleftmargin=0pt,
	xleftmargin=16pt,
	xrightmargin=3pt,
	comment=[l]{//},
	commentstyle={\color{darkgray}\ttfamily},
	emph={@software, @charging },
	emphstyle={\color{limegreen}},
	identifierstyle=\color{black},
	keywords={Given, When, Then, And},
	keywordstyle={\color{violet}\ttfamily},
	morecomment=[s]{/*}{*/},
	morestring=[b]",
	morestring=[s]{"""*}{*"""},
	ndkeywords={Scenario, Example, Feature},
	ndkeywordstyle={\color{darkviolet}\bfseries},
	sensitive=true,
	stringstyle={\color{javagreen}\ttfamily},
	numbers=none
}

\lstdefinestyle{JiraStyle} {language=Jira}

\title{Scenario-based Requirements Engineering for Complex Smart City Projects}

\author{
\IEEEauthorblockN{Carsten Wiecher\IEEEauthorrefmark{1}, 
				  Philipp Tendyra\IEEEauthorrefmark{1},
				  Carsten Wolff\IEEEauthorrefmark{1}}

\IEEEauthorblockA{\IEEEauthorrefmark{1}Dortmund University of Applied Sciences and Arts, IDiAL, Otto-Hahn-Str. 23, 44227 Dortmund\\
firstname.lastname@fh-dortmund.de, www.idial.institute
	}
}
\makeatletter
\def\ps@IEEEtitlepagestyle{%
  \def\@oddfoot{\mycopyrightnotice}%
  \def\@evenfoot{}%
}
\def\mycopyrightnotice{%
  {\begin{minipage}{\textwidth}
  \footnotesize \copyright 2022 IEEE. Personal use of this material is permitted. Permission from IEEE must be obtained for all other uses, in any current or future media, including reprinting\slash republishing this material for advertising or promotional purposes, creating new collective works, for resale or redistribution to servers or lists, or reuse of any copyrighted component of this work in other works.
  \end{minipage}
  }
  \gdef\mycopyrightnotice{}
}

\begin{document}
\maketitle


\begin{abstract}
Various stakeholders with different backgrounds are involved in Smart City projects. These stakeholders define the project goals, e.g., based on participative approaches, market research or innovation management processes.
To realize these goals often complex technical solutions must be designed and implemented.  
In practice, however, it is difficult to synchronize the technical design and implementation phase with the definition of moving Smart City goals. We hypothesize that this is due to a lack of a "common language" for the different stakeholder groups and the technical disciplines.
We address this problem with scenario-based requirements engineering techniques.
In particular, we use scenarios at different levels of abstraction and formalization that are connected end-to-end by appropriate methods and tools. This enables fast feedback loops to iteratively align technical requirements, stakeholder expectations, and Smart City goals. 
We demonstrate the applicability of our approach in a case study with different industry partners.  
\end{abstract}

\begin{IEEEkeywords}
Systems Engineering, Requirements Engineering, Project Management, Innovation Management
\end{IEEEkeywords}

\section{Introduction}
Concepts like Quality of Life (QoL) or Subjective Well-Being (SWB) can be used to derive expectations and goals for Smart City projects \cite{Ala-Mantila2018}; Systems engineering (SE) provides concepts for the development of complex socio-technical systems that address these goals~\cite{Dumitrescu2021}. However, in the interaction between the project environment and the SE processes, there is a lack of a "common language" to define, align, and continuously exchange requirements~\cite{Kasauli2021}. This leads to the risk that wrong system solutions are developed that do not meet the demand. 
Especially for socially-intense (soft) Smart City systems~\cite{Dustdar2017}.  
Therefore, deriving requirements for the SE project from stakeholder expectations is particularly difficult in projects with very diverse stakeholders, as it is the case for Smart City projects \cite{Turner2012}. 

We address this problem with the help of scenario-based requirements engineering (RE) techniques; Since scenarios are widely used in different disciplines \cite{Jarke1998}, we think that scenario-based techniques are suitable to interlink the stakeholder expectations (project environment) with the technical realization of products and services (project domain). 
This requires both the ability to specify scenarios in a generally understandable way for different stakeholders, and precise and unambiguous scenarios that are suitable as a basis for technical implementation \cite{Sutcliffe2003}.

To strengthen the interaction between the project environment and the project domain, we propose a new process (see Fig.~\,\ref{fig:toBeProcess}) together with appropriate tooling to support the specification, modeling, and management of scenarios throughout the system development process.
This process considers closed feedback loops to support an iterative alignment of
high level development goals and concrete technical requirements. 
We demonstrate the applicability of our process in a case study from an ongoing research project.\footnote{https://cilocharging.de}

In summary, we provide the following contributions~(C):
\begin{itemize}
    \item \textbf{C 1}: Based on our previous work \cite{Wolff2021}, we introduce a process for the scenario management and requirements validation with focusing on a Smart City context.   
    The process interlinks the project environment (the stakeholder view) with the technical systems development (the systems engineering view) and supports the continuous alignment of both views. 
    
    \item \textbf{C 2}: 
    We implemented the process. Therefore, we integrated the project management tool Jira\footnote{https://www.atlassian.com/software/jira}, with the requirements specification and modeling tool BeSoS \cite{Wiecher2021b}. 
    In this way, we can manage and specify requirements using scenario descriptions in natural language to support the stakeholder view. In addition, to support the systems engineering view, we formally model requirements using scenario-based modeling techniques \cite{Greenyer2021, Wiecher2020} supported by BeSoS (cf. \cite{Wiecher2021}).  
    Subsequently, with using the automation server~ Jenkins\footnote{https://www.jenkins.io}, we created a continuous requirements validation pipeline to automate the process, get fast feedback on newly added requirements, and hence create a strong link between the stakeholder and systems engineering view.  
    
    \item \textbf{C 3}: We show a case study from an ongoing research project. 
    The goal of the project is the development of  a  system  solution  for  city  logistics  with  electric vehicles.  
    In our case study we consider a system of systems (SoS) including several independent systems that are operated and managed by different industry partners in different organizations, where multiple stakeholders must be considered (e.g., grid operators, just-in-time logistics, consumers, traffic  planners, citizens and employees). Among other things (like applicability of the proposed method and tools), our case study demonstrates that our technique is suitable to specify scenarios at different levels of abstraction, and that low-level system scenarios are indeed helpful to align high-level SoS scenarios between stakeholders and different business cases. 
\end{itemize} 
This paper includes a summary of the addressed problem (Sect.~\,\ref{sect:problem}), an introduction of the proposed process (Sect.~\,\ref{sect:process}), an overview how scenarios-based requirements engineering techniques are applied within the process (Sect.~\,\ref{sect:processImplementation}), and a summary of the conducted case study (Sect.~\,\ref{sect:casestudy}). The paper ends with a lessons learned review, a summary and an outlook on future work. 

\section{State of the Art and Problem Statement}
\label{sect:problem}
As elaborated in \cite{Wolff2021}, our core hypothesis is that the SE processes and the project environment are not sufficiently interlinked within complex Smart City projects.
Consequently, this increases the risk that the results of the engineering activities and eventually the project outcome will not meet the expectations of the stakeholders \cite{Turner2012}; The outcome of the SE projects may create the expected value for the company delivering the product, but the individual solutions may not have the impact that stakeholders expect in a Smart City context. Accordingly, the question is how to best support the specification and alignment of requirements to address higher-level project goals (e.g., based on QoL, SWB concepts \cite{Ala-Mantila2018}) with concrete system requirements, e.g., for route planning systems, electric vehicles, or charging infrastructure. 

Recent studies have shown that a bottom-up approach (technology push) is often considered for the development of new systems for Smart Cities \cite{Daneva2018}; From the RE perspective, the focus is currently on specifying requirements that support the adaptation of existing systems for a Smart City context \cite{Daneva2018}. 
For the successful development of complex Smart City systems, however, both the top-down stakeholder view as well as the bottom up systems engineering view must be considered. It is necessary to \emph{a) elaborate the possible scenarios to reach higher level Smart City goals} (e.g., reduce traffic), and to \emph{b) investigate which new systems or changes in existing systems are necessary to reach these goals}. In addition, scenarios at both hierarchy levels must be consistently linked in order to \emph{c) identify profitable solutions for individual companies when focusing on the Smart City goals and, in the process, resolve possible 
conflicts between systems and companies}. 

This is challenging when adhering to traditional SE processes that follow clear development goals and focus on single systems and their subsystems \cite{Cavalcante2017}. 
Instead, a system of systems engineering (SoSE) perspective is more suitable when specifying requirements for Smart City projects \cite{Nielsen2015, Cavalcante2017}. However, especially for the specification, modeling and management of requirements for SoS, tools and methods are missing \cite{Ncube2018}.

\section{Process Overview}
\label{sect:process}
To address these problems, we propose a process as shown in Fig.~\,\ref{fig:toBeProcess}. As already introduced in \cite{Wolff2021}, this process addresses two core topics: 
\begin{figure}[h]
    \centering
    \includegraphics[width=1.0\linewidth]{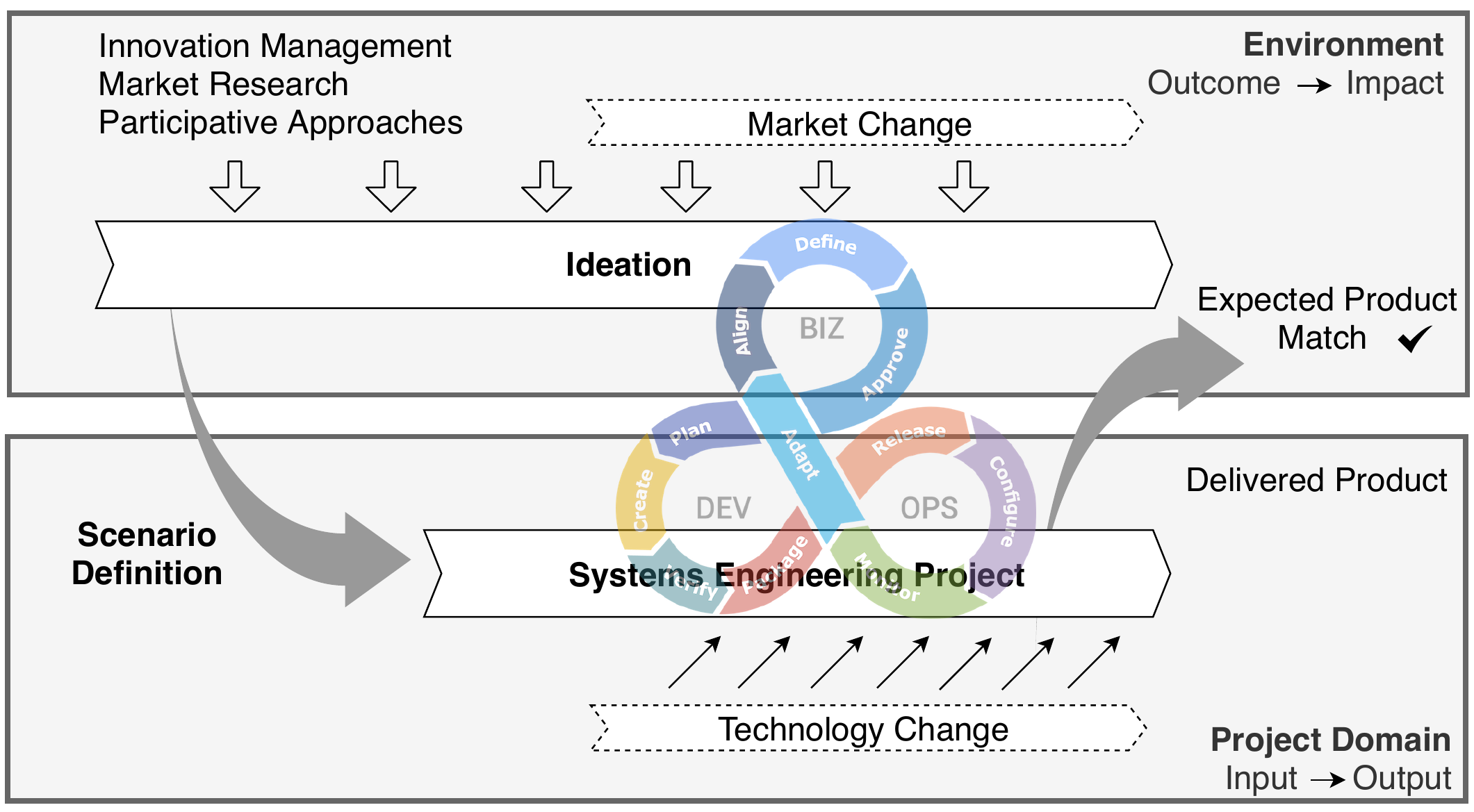}
    \caption{To-be-process for complex systems engineering projects \cite{Wolff2021}.}
    \label{fig:toBeProcess}
\end{figure}

    \emph{1) Continuous interaction between the ideation phase and the SE project:} As an interface to the project environment, the \emph{ideation phase} is carried out continuously until the end of the SE project. In this way, innovation management, market research, or other techniques continuously provide new information for the SE project. This information is continuously collected within the \emph{scenario definition} step. The main difference to traditional approaches is that we do not consider the ideation phase and the SE project as independent and separated activities. Instead, we use the scenario definition step to iteratively add new information in small increments based changes in the project environment. 

    \emph{2) Early system prototypes \& BizDevOps}: We use the \emph{BizDevOps} approach to support a continuous update of requirements. Thereby, the project environment (Biz) is supported in defining scenarios that describe expectations regarding the project outcome and the possible impact. Subsequently, we use these high level scenarios to derive concrete and executable system scenarios (Dev). These executable system scenarios are used as early prototypes of the system in development that can be validated and aligned with the project environment~(Ops)~(cf. \cite{Wiecher2020}). As a result of the scenario definition we create two interlinked artifacts: backlog issues as input for the project management activities in the SE project, and a formal and validated scenario specification as input for the concrete systems design. 

\section{Process Implementation}
\label{sect:processImplementation}
\begin{figure*}
    \centering
    \includegraphics[width=1.0\linewidth]{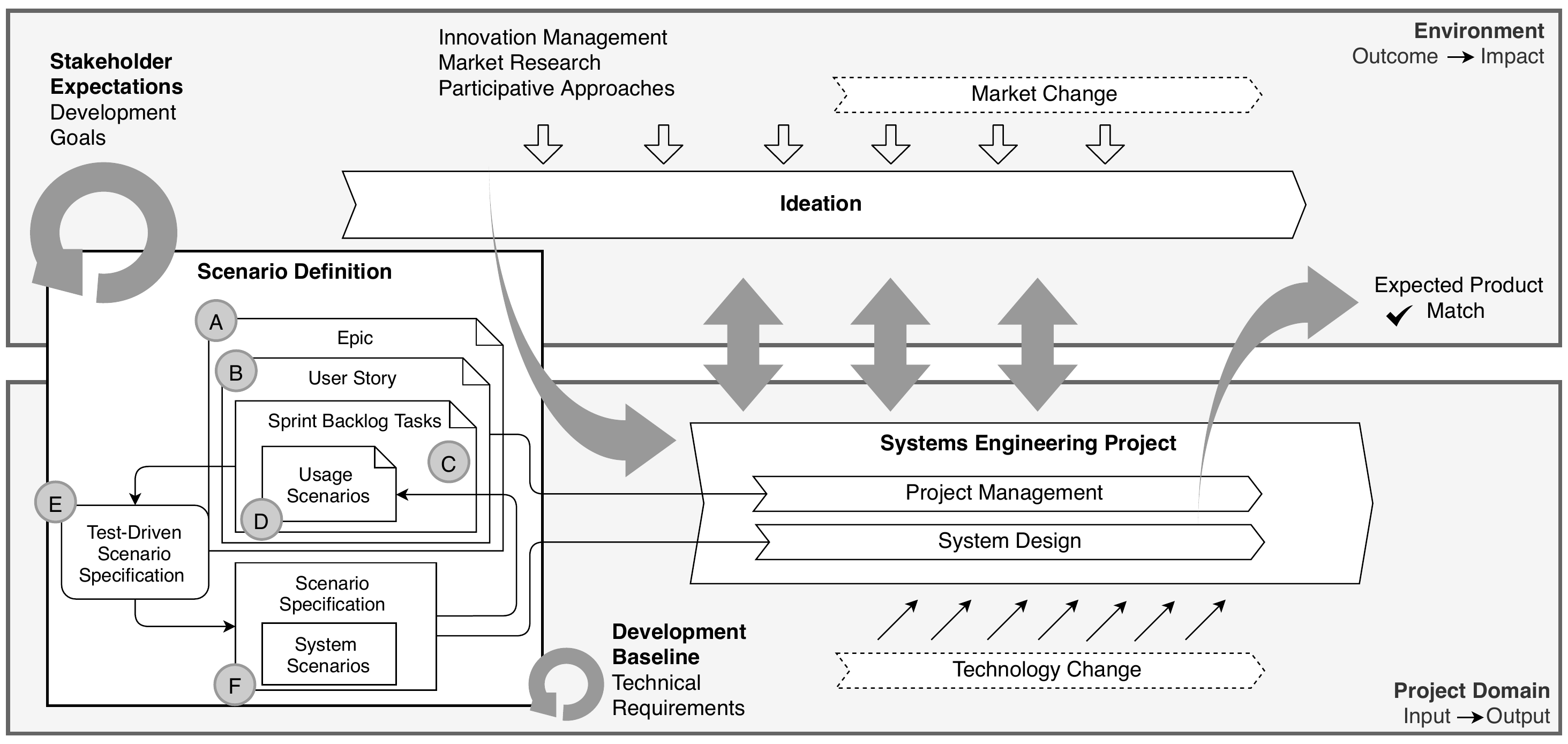}
    \caption{To-be-process including the implementation of the scenario definition.}
    \label{fig:process}
\end{figure*}
To realize the process concept, we propose an implementation as outlined in Fig.~\,\ref{fig:process}, which we illustrate by going through the single steps of the scenario definition.  

\subsection{Project Environment and Smart City Goals}
To realize the interface to the project environment, we use agile project management artifacts to systematically collect and manage Smart City goals. As an entry point we propose to use \emph{epics} (see A in Fig. \ref{fig:process}) to define high level scenarios as shown in Listing~\,\ref{list:epic}. 
\begin{lstlisting}[caption=Epic to formulate a Smart City goal,
	label=list:epic,
	style=JiraStyle
	]
As a logistics service provider I expect that the current 
charging status and location of the vehicles are taken into
account when planning routes, to reduce the energy costs of 
the vehicle fleet. 
\end{lstlisting}

Thereby, to manage the various goals of different stakeholders, the single epics are annotated with \emph{tags} that indicate the relationship between epics and Smart City goals. E.g., one group of epics can be linked to \emph{Goal1: Reduce pollution cased by inner-city delivery traffic}, and another group of epics can be linked to \emph{Goal2: Equalize the energy consumption in the city district.}

\subsection{Transition to SE Project Management} 
Subsequently, we use the defined epics to create concrete \emph{user storys}~(B) that document specific features of a system from the perspective of users and stakeholders, as exemplified in Listing~\,\ref{list:story}. 
\begin{lstlisting}[caption=User story to concretize an epic,
	label=list:story,
	style=JiraStyle
	]
As a delivery vehicle driver, I want to have a list of 
optimized routes to shorten delivery time. 
\end{lstlisting}

Consequently, the resulting documentation is a good indication of the value each feature can create for a specific stakeholder. 
In doing so, the epics and user stories support the integration of the project environment (Biz). In addition, the user stories are the entry point for the specification of concrete system scenarios in order to provide technical requirements for the system design.  

\subsection{Technical Requirements for the SE Project}
To create a strong link between epics, the contained user stories (Biz), and technical requirements (Dev), we integrated the project management tool Jira with the scenario modeling tool BeSoS~\cite{Wiecher2021b}. 
Methodically, this is done by applying the Behavior-Driven Development (BDD) technique on a requirements specification level. For this purpose, we extend the stakeholder-oriented user stories with \emph{usage scenarios} (D) that describe how a concrete interaction with the system takes place. Technically this is done using the gherkin syntax as part of Cucumber\footnote{https://cucumber.io} features (C). E.g., based on the user story in Listing~\ref{list:story}, we can define a usage scenario as shown in Listing \ref{list:gherkin}. 
\begin{lstlisting}[caption=Usage scenario to concretize user story of Listing \ref{list:story},
	label=list:gherkin,
	style=GherkinStyle
	]
Feature: calculate optimized routes for delivery vehicle driver

  Scenario: the vehicle driver request new routes after the vehicle was loaded with freight
    When the vehicle driver request routes with the navigation system of the vehicle 
    Then the navigation system forwards information on the loaded freight to the route planning system 
    And the route planning systems calculates different route options 
\end{lstlisting}

As a next step, following the Test-Driven Scenario Specification (TDSS) approach (E) \cite{Wiecher2020}, we generate test steps from usage scenarios. These tests are used to drive the modeling of technical requirements. In this way we iteratively execute the generated test steps and add scenarios to a \emph{scenario specification} (F) until all tests pass.\footnote{For brevity, we refer to our previous work that includes technical details on the BDD specification method for SoS \cite{Wiecher2021}.}     
This TDSS approach has the advantage that we focus on the overall Smart City goals when modeling the technical requirements, as the test steps are generated from usage scenarios associated with user stories, epics, and goals within the project management tool. 
Consequently, we create a continuous traceability from goals to technical system requirements. 

The concrete modeling of technical requirements has the form as shown in Listing~\,\ref{list:scenarioSpecification}. This Listing shows a scenario representing a requirement that realizes the usage scenario of Listing~\,\ref{list:gherkin}. A scenario specification includes several of these scenarios that, when executed, exchange information by events that can be requested by a scenario or that a scenario can wait for.\footnote{For brevity, we omit a detailed description of the modeling technique here. For a description of the basic concepts please see \cite{Greenyer2021}. For their application, see the BeSoS introduction \cite{Wiecher2021b}.} E.g., according to the usage scenario of Listing~\,\ref{list:gherkin}, the modeled requirement in Listing~\,\ref{list:scenarioSpecification} will be executed when the vehicle driver requests routes (trigger event in line 1, \lstinlineKotlin{When} step in Listing~\,\ref{list:gherkin}). Subsequently, in the body of the scenario (line 2 and 3), we request events that describe the expected system behavior. In this example, the navigation system forwards information to the route planner (\lstinlineKotlin{Then}), and the route planning system calculates different route options~(\lstinlineKotlin{And}). 
\begin{lstlisting}[caption=Scenario specification,
	label=list:scenarioSpecification,
	style=KotlinStyle
	]
scenario(vehicleDriver sends navi.requestRoute()){
    request(navi sends routePlaner.freightInfo())
    request(routePlaner.calculateRoutes())
}
\end{lstlisting}

This way of modeling technical requirements has the advantage that scenarios can be added and updated iteratively according to changing project environment requests. And, by applying TDSS, we execute and automatically analyze the scenario specification. In this way we can automatically detect contradictions in requirements (see e.g. \cite{Wiecher2020}). This means that, e.g., two stakeholders may request changes in two different systems that are only loosely coupled at the SoS level. In this case, it is difficult to identify possible contradictions between these two changes when the requirements are analyzed manually. 
The proposed iterative specification, modeling, and automated analysis of requirements (cf. \cite{Wiecher2021b, Wiecher2021}) is therefore a 
significant benefit for complex Smart City projects, as it allows working with small increments and focusing on concrete new solutions, while the automated analysis of the entire scenario specification shows whether the new increments lead to inconsistencies in the specification.

\subsection{Automation}
\begin{figure}[h]
    \centering
    \includegraphics[width=1.0\linewidth]{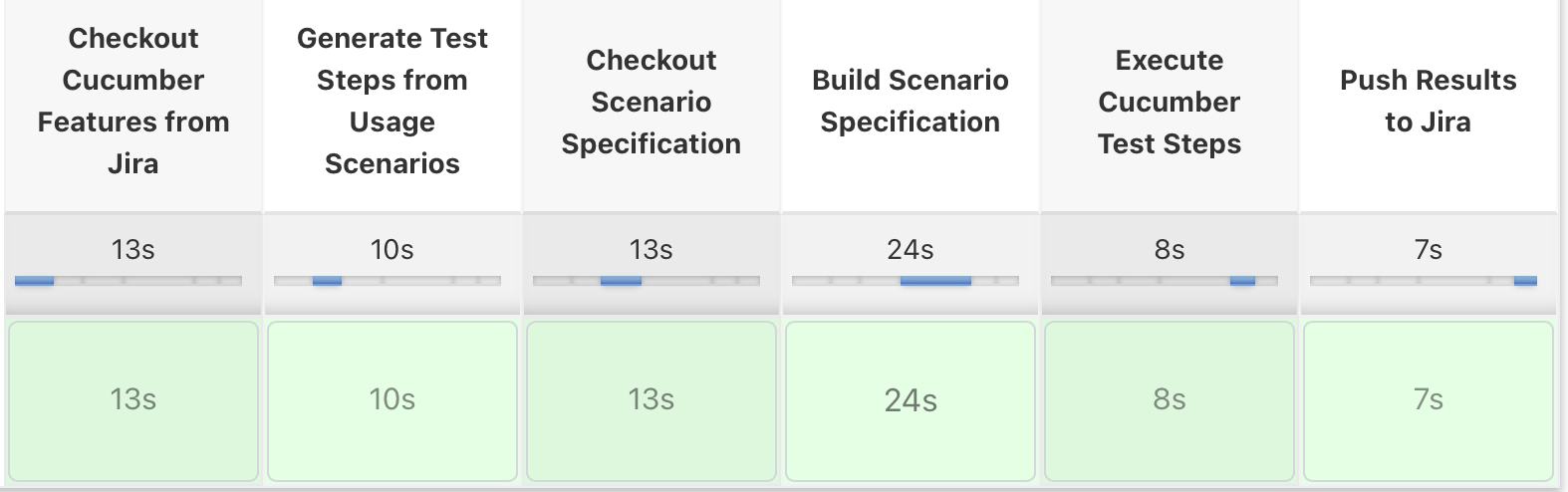}
    \caption{Requirements validation pipeline to align technical requirements with the living documentation in Jira.}
    \label{fig:jenkins}
\end{figure}

For a successfully application of our approach, we propose an automated and continuous pipeline as shown in Fig.~\,\ref{fig:jenkins}. 
This pipeline is realized with the automation server Jenkins and directly links the single artifacts. The execution of the pipeline is triggered by two sources. First, when the project environment provides new information and the feature specification (C in Fig.~\,\ref{fig:process}) is changed (Biz). Second, when the scenario specification (F) in the source code repository is updated~(Dev), i.e. new requirements were modeled or the existing implementation was changed.  
In both cases, the complete pipeline is executed to: 1) checkout the feature specification from Jira, 2) generate new test steps from the containing usage scenarios, 3) checkout the scenario specification from the repository, 4) compile the specification, 5) execute the test steps, and eventually 6) push  the results back to the Jira project. In this way we get direct feedback, if, e.g., a feature does not have a corresponding technical requirement, or conflicts between requirements and systems occur while updating the scenario specification.  

\section{Case Study}
\label{sect:casestudy}
To asses the feasibility of our approach, we conducted a case study from the CiLoCharging project.\footnote{https://cilocharging.de}
This project develops a system solution for city logistics with electric vehicles. 
Such vehicles need to be charged with electrical energy and loaded with goods in a logistics depot. Therefore, this depot is a relevant consumer of electricity in the smart grid and a relevant source of traffic in the city’s traffic management. In addition, it can control and manage the energy demand and the traffic by optimizing the charging and the logistics processes. Ultimately, the depot and the electric vehicles become active, intelligent components in the Smart City. Multiple stakeholder perspectives have to be integrated, e.g. grid operators, just-in-time logistics, consumers, traffic planners, and citizens with their wish for timely parcel delivery with minimum disturbance by traffic. 
\begin{figure*}[h]
    \centering
    \includegraphics[width=1.0\linewidth]{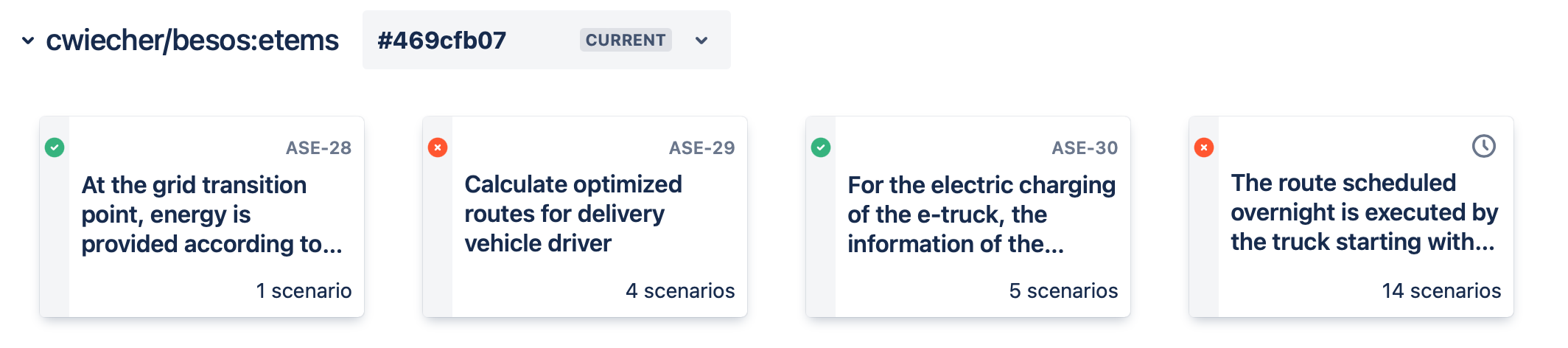}
    \caption{Living requirements documentation in Jira.}
    \label{fig:documenetation}
\end{figure*}

\begin{figure}[h]
    \centering
    \includegraphics[width=1.0\linewidth]{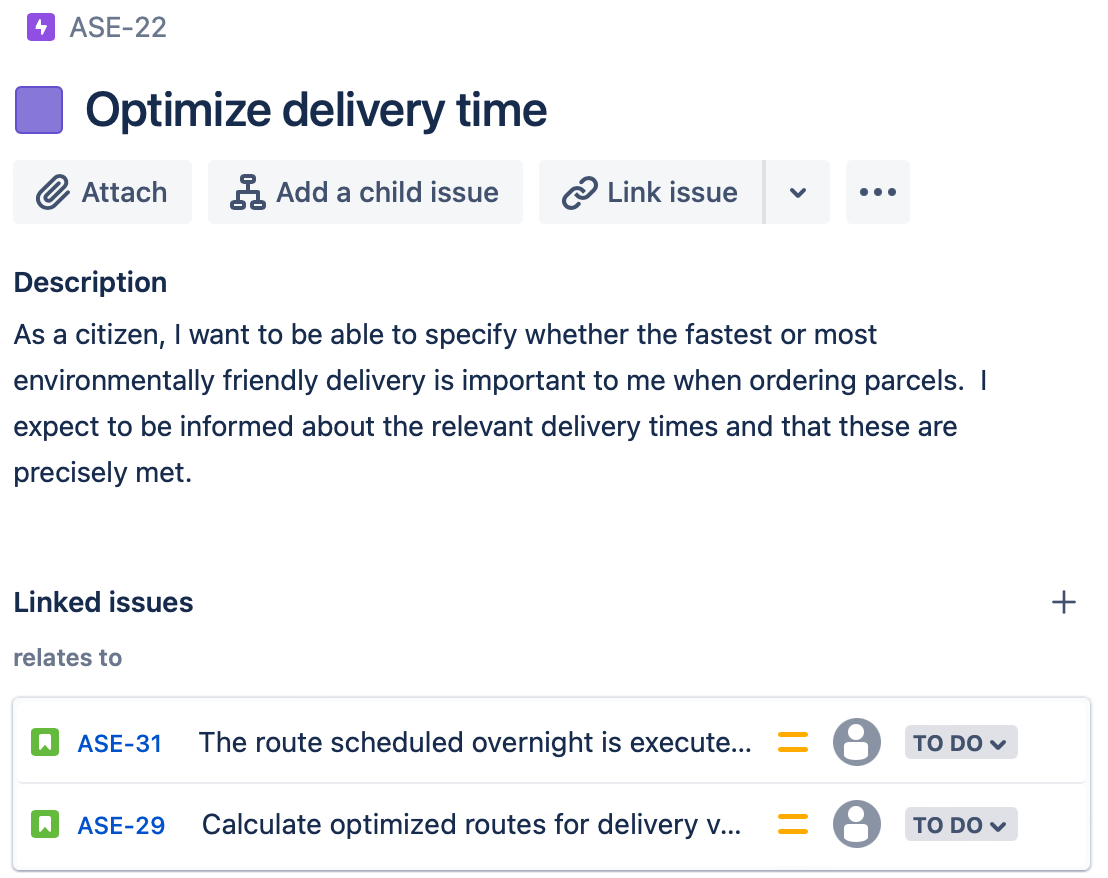}
    \caption{Epic description including linked user storys.}
    \label{fig:epic}
\end{figure}

Based on input from our industry partners, we executed several iterations of the scenario definition part of our process. We first started with a top-down perspective by identifying Smart City goals and related epics that we documented in Jira. One example is shown in Fig.~\,\ref{fig:epic}. Here we see an epic that documents a high level requirement from the perspective of a citizen. 
Within the project context, this requirement could not be realized by a single industry partner. Instead, we identified several systems (driver, route planner, traffic and logistics management) that must be integrated to realize the stakeholder request. 
Consequently, in a second step, we added user storys to the epics to define concrete requirements for the single systems and companies (see e.g. ASE-31 and ASE-29 in Fig.~\,\ref{fig:epic}) that were necessary to realize the epic.    

In this way, we were able to systematically define high level requirements and specify user storys that must be implemented. Both steps support the stakeholder view. To learn if the transition to the systems engineering view is also applicable within the project context, we started to refine 
each user story by specifying concrete user and system interactions. Therefor, we added usage scenarios to each user story within the BeSoS tool. Since we integrated the high level documentation in Jira with the specification of system interactions in BeSoS, we were able to create a consistent "living documentation" as shown in Fig.~\,\ref{fig:documenetation}. Here we see Cucumber features that include the identified usage scenarios, and each feature is linked to the related user story (e.g., ASE-29 that was specified to realize the epic ASE-22). 

\begin{figure}[h]
    \centering
    \includegraphics[width=1.0\linewidth]{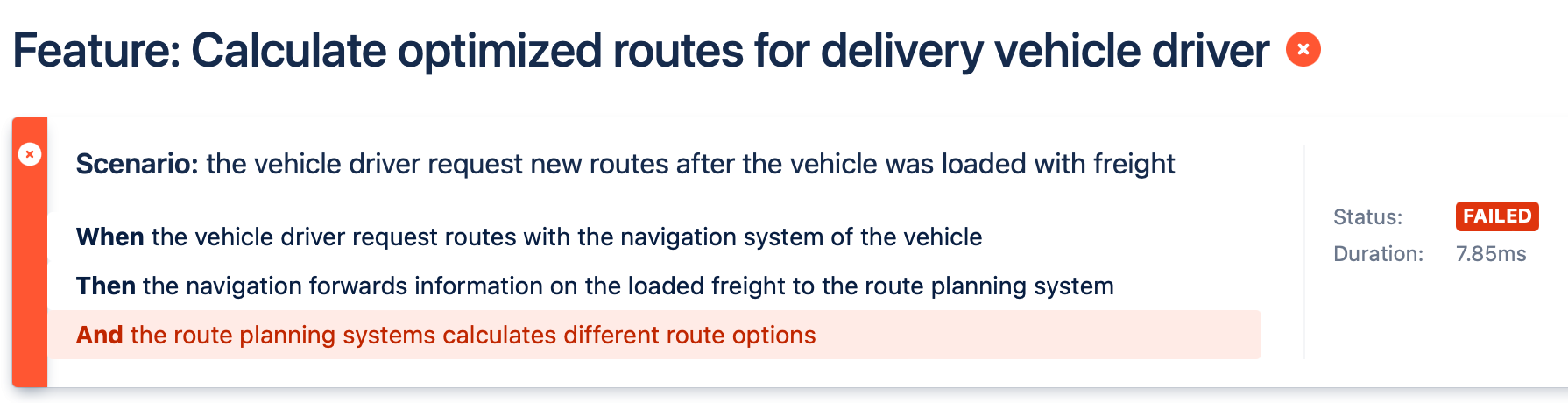}
    \caption{Failed feature test}
    \label{fig:failedTest}
\end{figure}

At this stage we also used the proposed requirements validation pipeline (see Fig. \ref{fig:jenkins}) in order to utilize this "living documentation" as the BizDevOps synchronization point. In this way, it was not only possible to create a consistent documentation of stakeholder expectations and requirements, but it was also possible to use this documentation as a trigger for the automated validation of the technical requirements, and to read the validation results back into the documentation. As an example, the usage scenario in Fig. \ref{fig:failedTest} is part of a feature that is linked to the user story ASE-29. As we started with the top-down perspective, there was no modeled technical requirement in the form of Listing \ref{list:scenarioSpecification}. Consequently, when the automation server executed the feature, we did not receive the expected system response and, accordingly, got a failed test result. 

\begin{figure}[h]
    \centering
    \includegraphics[width=1.0\linewidth]{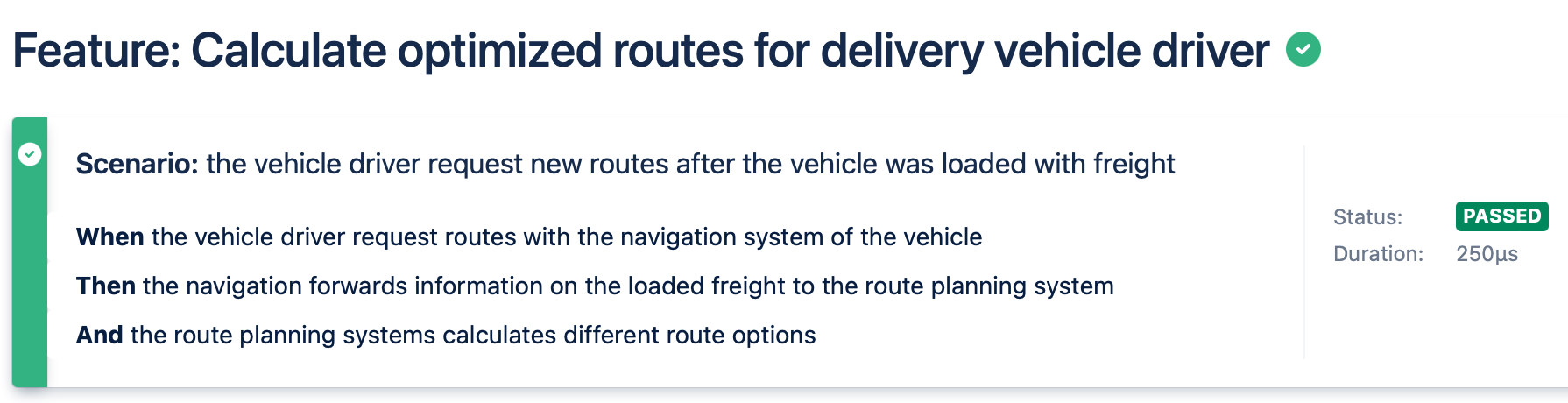}
    \caption{Passed feature test}
    \label{fig:passedTest}
\end{figure}

To correct this test result, we implemented the technical requirement, and after it was added to the repository, the validation pipeline was triggered again, resulting in the updated documentation, as shown in Fig. \ref{fig:passedTest}. 

\section{Lessons Learned}
\label{sect:lessonslearned}
While performing the individual steps of the scenario definition within the CiLoCharging project, we gained several insights, of which we will briefly describe the most important ones: 
\begin{itemize}
    \item The use of epics and user storys is useful to concretize Smart City goals: In the project context with the different industry partners, the way to specify epics for specific Smart City goals was useful to document a SoS perspective. In this way, it was possible, for example, to define SoS requirements from a citizen's perspective and link user storys that contribute to those SoS requirements.    
    \item The concretization of user storys with usage scenarios is practical: The concretization of user storys by usage scenarios is a pragmatic approach to iteratively document the expected system behavior. When starting top-down with Smart City goals, it was possible to directly formulate the expected system behavior as a usage scenario. However, it was difficult to decide how many usage scenarios are needed until a system behavior is sufficiently specified. 
    \item Scenario-based modeling of technical requirements is suitable: Our way of specifying usage scenarios to drive the scenarios-based modeling of technical requirements is an appropriate combination, since the expected behavior can directly be  translated into an executable system scenario. However, this manual translation requires additional modeling effort and knowledge. 
    \item Automation of requirements validation supports the synchronization of the stakeholder and SE view: By modeling technical requirements driven by tests that were systematically derived from the stakeholder view, it was possible to create and validate the end-to-end chain of requirements. When automatically synchronized, the resulting "living documentation" creates a strong link between the two different views.  
\end{itemize}

\section{Closing and Next Steps}
In our previous work \cite{Wolff2021} we outlined a new process to improve the interaction between the project environment and the project domain in complex Smart City projects. This paper shows a first implementation with focusing on the scenario definition part. We applied our implementation in a first case study. Although we used and integrated state-of-the-art tools and chose a realistic Smart City scenario to validate our process, the use case only considers a small data set compared to real-world scenarios. Next steps will focus on larger data sets, and we will integrate techniques to support the prioritization of requirements. In addition, we will connect our approach with artifacts within SE projects
\cite{Wiecher2021c} to increase the applicability in real world projects. 

\bibliographystyle{IEEEtran}
\bibliography{references}

\end{document}